\documentclass{arXiv}
\usepackage{threeparttable}
\usepackage{natbib}
\usepackage{pdflscape}
\usepackage{graphicx}

\begin{document}
\SetRunningHead{R. A. Burns et al.}{Distance to IRAS 20056+3350: MSFR on the Solar circle}

\title{Trigonometric Distance and Proper Motion of IRAS 20056+3350: Massive Star Forming Region on the Solar Circle}

\author{Ross A.   \textsc{Burns},\altaffilmark{1}
        Takumi   \textsc{Nagayama},\altaffilmark{2}
        Toshihiro \textsc{Handa},\altaffilmark{1}
        Toshihiro \textsc{Omodaka},\altaffilmark{1}
        Akiharu   \textsc{Nakagawa},\altaffilmark{1}
        Hiroyuki \textsc{Nakanishi},\altaffilmark{1}
        Masahiko   \textsc{Hayashi},\altaffilmark{3,4}
        Makoto   \textsc{Shizugami},\altaffilmark{5}
        }
\altaffiltext{1}{Graduate School of Science and Engineering, Kagoshima University,\\
                 1-21-35 K\^orimoto, Kagoshima, Kagoshima 890-0065, Japan}
\email{RossBurns88@MilkyWay.sci.Kagoshima-u.ac.jp}
\altaffiltext{2}{Mizusawa VLBI Observatory, National Astronomical Observatory of Japan, \\
                 2-21-1 Osawa, Mitaka, Tokyo 181-8588, Japan}
\altaffiltext{3}{National Astronomical Observatory of Japan, \\
                 2-21-1 Osawa, Mitaka, Tokyo 181-8588, Japan}
\altaffiltext{4}{School of Mathematical and Physical Science, The Graduate University for Advanced Studies \\
                (SOKENDAI), Hayama, Kanagawa 240-0193, Japan}
\altaffiltext{5}{Mizusawa VLBI Observatory, National Astronomical Observatory of Japan, \\
                 2-12 Hoshi-ga-oka, Mizusawa-ku, Oshu, Iwate 023-0861, Japan}

\KeyWords{Masers - Stars; individual (IRAS 20056+3350) - Galaxy: structure and dynamics} 

\maketitle

\begin{abstract}
We report our measurement of the trigonometric distance and proper motion of IRAS 20056+3350, obtained from the annual parallax of H$_{2}$O masers. 
Our distance of $D = 4.69^{+0.65}_{-0.51}$ kpc, which is 2.8 times larger than the near kinematic distance adopted in the literature, places IRAS 20056+3350 at the leading tip of the Local arm and proximal to the Solar circle.
Using our distance we re-evaluate past observations to reveal IRAS 20056+3350 as a site of massive star formation at a young stage of evolution. This result is consistent with the spectral energy distribution of the source evaluated with published photometric data from UKIDSS, WISE, AKARI, IRAS and sub-millimetre continuum. Both analytical approaches reveal the luminosity of the region to be $2.4\times 10^{4} L_{\odot}$, and suggest that IRAS 20056+3350 is forming an embedded star of $\geq 16 $M$_{\odot}$.
We estimated the proper motion of IRAS 20056+3350 to be ($\mu_{\alpha}\cos\delta$, $\mu_{\delta}$) = ($-2.62\pm0.33$, $-5.65\pm0.52$) mas yr$^{-1}$ from the group motion of H$_{2}$O masers, and use our results to estimate the angular velocity of Galactic rotation at the Galactocentric distance of the Sun, $\Omega_{0} = 29.75\pm2.29$ km s$^{-1}$ kpc$^{-1}$, which is consistent with the values obtained for other tangent point and Solar circle objects.

\end{abstract}

\newpage

\section{Introduction}

The morphology of a galaxy is determined visually and is highly influenced by the distribution of massive stars which are bright and typically confined to the spiral arms \citep{Uq14,Reid14}. 
Although the exclusivity of massive star formation to the spiral arms is not yet understood it does bring to light a strong interplay between the spiral pattern itself and the conditions required to trigger massive star formation. Thus, neither phenomena can be understood independently; we cannot explain massive star formation without explaining the role of the spiral pattern and we require the by-products of massive star formation to understand the morphology of galaxies by tracing the arms.

From our viewpoint within the Galactic disk, perhaps the most important parameter common to both fields of investigation is distance. In addition to the clear necessity of accurate distances for mapping the Galactic arms, the interpretation of the physical properties of a massive star forming region (MSFR) is greatly influenced by the distance estimate adopted - since many of these parameters have a non-linear relationship with distance. 

For convenience, many investigators make use of the kinematic distances of Galactic star forming regions (SFRs). However, kinematic distance calculations require pre-calibration of the Galactic rotation curve and suffer from distance ambiguities ($D_{\rm near}$ and $D_{\rm far}$) for objects in the inner Galaxy. Moreover, they are unreliable when the source has a large peculiar velocity with respect to purely circular rotation in the Galactic disk. Consequently, the kinematic distance can be poorly determined for objects in some particular locations such as in the direction of the Galactic center and anticenter, and Local Arm. 

Measurements of trigonometric parallaxes to Galactic MSFRs directly improves the reliability of their distances. Such experiments involve measurement of the annual parallax of MSFRs typically via astrometric monitoring of maser spots using VLBI networks such as the VLBI Exploration of Radio Astrometry (VERA), the Very Long Baseline Array (VLBA), and the European VLBI Network (EVN). In addition to distance measurement, masers are also extremely useful for understanding the processes of star formation itself since masers emitted by different molecular species and transitions are known to be associated with different physical environments \citep{Bart12}. As a result, masers are often seen to trace structures such as expanding shells and shock fronts \citep{Trinidad13}, and bipolar outflows (\citealt{Imai07,NY08,Moscadelli11,Torrelles14} and G236.81+1.98 in \citealt{Choi14}). Masers thus allow the internal motions of star formation mechanisms to be seen even though the structures themselves may not be directly unobservable due to the embedded nature of such systems.

Finally, combining the proper motions and line of sight motions of masers sometimes allows estimation of the three-dimensional secular motions of a star forming region in the plane of the Galaxy. This is one of the unique advantages of VLBI maser investigations and is a dominant approach to understanding the kinematics and structure of the Milky Way Galaxy (MWG) via the evaluation of the Galactic constants, $R_{0}$, $\Theta_{0}$ and $\Omega_{0}$ \citep{Honma12,Reid14}. These parameters can be evaluated more reliably for SFRs that reside at special locations in the MWG such as the tangent points \citep{Nagayama11a,Burns14} and Solar circle \citep{Ando11}. 


All VLBI observations discussed in this paper were carried out using VERA \citep{Koba03} which is a Japanese VLBI array dedicated to measuring the annual parallax of maser sources in the MWG.
IRAS 20056+3350 was chosen for this investigation for its bright and stable maser emission at 22 GHz - first seen in \citet{Jenness95} - and its presumed proximity to the Solar circle. IRAS 20056+3350 is listed in the IRAS catalogue of point sources \citep{IRAS88}.

In this work we aim to make a contribution to each of the aforementioned topics. This paper continues as follows: Observations and data reduction are discussed in \S2. Results are reported in \S3, including analyses of the astrometric accuracy achieved in parallax fitting of maser spots. The physical nature of the IRAS 20056+3350 MSFR is explored in \S4 in the context of archive and publicly available data, and its location is discussed with respect to other MSFRs who's distances have also been measured with VLBI annual parallax. Finally in this section we evaluate $\Omega_{0}$ using the results obtained from our observations with VERA. Conclusions are reported in \S5.\\


\section{Observations and data reduction}
Data were obtained using VERA in dual-beam mode. By observing H$_{2}$O masers in IRAS 20056+3350 and the J2010+3322 reference continuum source simultaneously we calibrated tropospheric phase fluctuations using the reference source and applied solutions directly to the maser data in real-time without interpolation. The scan integration time of the pair was about 9 minutes. Intermittent observations of BL Lac or 3C454.3 were made every 1.5 hours for bandpass calibration. A typical observation session lasted roughly 8 hours, providing $\sim$ 3.3 hours on-source integration and good coverage in the uv plane. 
Phase tracking centers for the maser and continuum source were set to
$(\alpha, \delta)_{\mathrm{J}2000.0}=(20^{\mathrm{h}}07^{\mathrm{m}}31^{\mathrm{s}}.2593$,
+33$^{\circ}$59'41".491) and 
$(\alpha, \delta)_{\mathrm{J}2000.0}=(20^{\mathrm{h}}10^{\mathrm{m}}49^{\mathrm{s}}.7233$,
+33$^{\circ}$22'13".810), respectively.
The positional reference J2010+3322 is listed in the VLBA calibrator search catalogue \citet{VLBA2}.

Left-handed circular polarisation signals were sampled with 2-bit quantisation, and filtered with the VERA digital filter unit \citep{Iguchi05}.
The total available bandwidth of 256 MHz was divided into 16 intermittent frequency (IF) channels, each with a bandwidth of 16 MHz. One IF was centered on the maser emission, assuming a rest frequency of 22.235080 GHz, and the other 15 IFs, in adjoining frequency, were allocated to the continuum reference source. 

Seven epochs of observations were carried out over a period slightly short of two years with epochs typically spaced 3-6 months apart. A summary of the observing calendar is given in Table~\ref{table:1}. Maser spots were readily identifiable from one epoch to another due to the stability of emission and the low number of maser features (\emph{see Section 3.1}).

\null

\begin{table}[!h]

\caption{Summary of observations made with VERA.\label{table:1}}
\begin{center}
\small
\begin{tabular}{cccccc}
\hline
Observation&&Detected\\
Epoch&Date&spots\\ \hline
1&2012 Feb 05&2\\
2&2012 Apr 30&2\\
3&2012 Aug 09&2\\
4&2013 Feb 16&3\\
5&2013 Apr 22&3\\
6&2013 May 05&2\\
7&2013 Dec 24&2\\
\hline
\end{tabular}

\end{center}

\end{table}

\begin{table*}[]
\hspace*{+1.2cm}
\begin{threeparttable}[c]
\caption{The general properties of H$_{2}$O maser in IRAS 20056+3350 detected with VERA.\label{table:2}}
\begin{center}
\small
\begin{tabular}{ccccccccc}
\hline
Spot&$V_{\rm LSR}$&Detected &$\Delta \alpha \cos \delta$&$\Delta \delta$&$\pi$ &$\mu_{\alpha}\cos\delta$&$\mu_{\delta}$ \\ 
ID&(km s$^{-1}$)&epochs&(mas)&(mas)&(mas)&(mas yr$^{-1}$)&(mas yr$^{-1}$)\\ \hline
A&$+7.27$&1234567&$0$&$0$&$0.213\pm 0.028$&$-2.62\pm0.14$&$-6.04\pm0.03$\\ 
B&$+7.27$&12345**&$-2.03$&$+0.65$&$0.212\pm0.055$&$-3.33\pm0.23$&$-6.00\pm0.05$\\ 
C&$-2.25$&****456*&$-521.69$&$+581.19$&&$-1.92\pm0.14$&$-4.91\pm0.03$\\ 
D&$+0.25$&*******7&+804.61&+3.47&&&\\
&&&Group fitting&&$0.213 \pm 0.026$&&&\\ \hline  
&Average &&&&&$-2.62 \pm 0.33$&$-5.65 \pm 0.52$&\\
\hline
\end{tabular}
\begin{tablenotes}
\item{Column (3)}: Epoch numbers assigned in column(1) of Table 1, while asterisk represents non-detection.
\end{tablenotes}
\end{center}
\end{threeparttable}
\end{table*}

Interferrometric correlation was carried out using the Mitaka FX correlator \citep{Chikada}. A frequency resolution of 15.625 kHz,
corresponding to a velocity resolution of 0.21 km s$^{-1}$, was used for the maser data, and for the continuum source a frequency resolution of 250 kHz was used.
We did not use the central IF of the continuum source data during data reduction since the frequency resolution of this IF channel did not match that of the other 14 IFs. This is an affect of the correlation process whereby the frequency resolution of this IF reflects that of the maser data. The small loss of bandwidth does not effect our goals adversely since the continuum source is readily detected with a signal to noise ratio of $\geq 80$ without the inclusion of the central IF channel.

GPS measurements of the atmospheric water vapour content at each station were used to refine an a-priori model of tropospheric delay expected during the rise and fall of source elevation \citep{Honma08b}. Model solutions were applied post-correlation. In the 6th epoch, GPS data were not available at VERA's Ogasawara station. For this epoch we instead used water vapour measurements from the Japan meteorological association (JMA) to the same effect, though at a lower time resolution. The reliability of these procedures is explored in detail in Nagayama (2014, in preparation).

All data were reduced using Astronomical Image Processing System (AIPS) developed by the National Radio Astronomy Observatory (NRAO).
Amplitude and bandpass calibration for both beams was carried out using the standard procedures of reduction of astrometric VLBI data in AIPS. During the first round of data reduction we created phase referenced images of the maser emission by calibrating maser data using phase solutions obtained for J2010+3322 in the AIPS task FRING. The positions of masers were determined by the peaks of 2D Gaussian fits applied to the final images. Using this procedure we found the two brightest maser spots (spot ID:A and spot ID:B; \emph{see Table~\ref{table:2}~}). The other spots (spot ID:C and spot ID:D) were only seen in self-calibrated images created in the second round of data reduction. In this reduction procedure we made single-beam images by self-calibration using the AIPS task pair IMAGR and CALIB on the brightest maser spot (spot ID:A). Subsequent maser positions were determined relative to the astrometric position of the bright reference maser, which was determined from the phase-referenced map.

\section{Results}

\subsection{Distribution of H$_{2}$O masers in IRAS 20056+3350}
A modest total of 4 individual maser spots were detected in our observations. Maser spot detections are summarised in Table~\ref{table:2} where position offsets are given relative to the reference maser (Spot ID:A) for which we measure first epoch absolute co-ordinates of $(\alpha, \delta)_{\mathrm{J}2000.0}=(20^{\mathrm{h}}07^{\mathrm{m}}31^{\mathrm{s}}.2586$,
+33$^{\circ}$59'41".477). The LSR velocity domain of masers, ascertained from Doppler shifts, ranges from $-2.25$ to $+7.27$ km s$^{-1}$. Thus, all detected masers are blueshifted with respect to the parent cloud - whose velocity from molecular line observations is known to be $+9$ km s$^{-1}$ (\emph{see section 4.2.1}). 
The spatial distribution and velocities of maser spots detected in IRAS 20056+3350 are presented in Fig.~\ref{fig:4} with arrows indicating proper motion vectors (\emph{see section 3.4}).

\subsection{Annual parallax}
Parallax and proper motion fitting was performed simultaneously on the two maser spots which were identifiable in multiple observations, spanning more than one year. Astrometric motions of these spots were deconstructed into linear and sinusoidal components arising from the sky-plane proper motion and annual parallax, respectively. In the fitting procedure, nominal fitting errors in the R.A. and Dec. directions were applied and reduced itteratively until a $\chi^{2}$ value of unity was reached. These error floors were 0.227 mas in R.A. and 0.050 mas in Dec.

\begin{figure}[h!]
\begin{center}
\hspace{-0.33cm}
\includegraphics[scale=0.8]{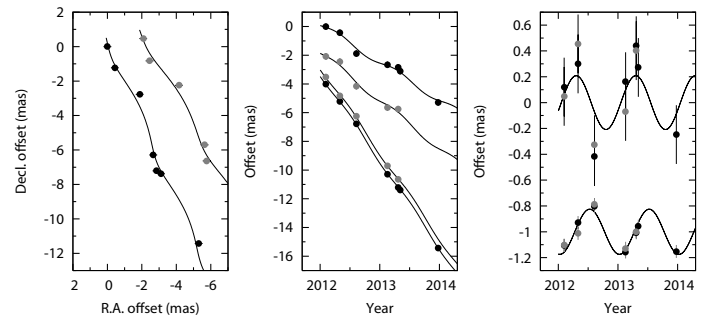}
\caption{Parallax and proper motion fitting for two maser spots in IRAS 20056+3350. Spot ID:A and ID:B are represented by black and grey points respectively. Left: Sky-plane motion of the maser spots. Middle: motion in R.A. (above) and Dec. (below) as a function of time. Right: parallactic motion in R.A. (above) and Dec. (below) after subtraction of the linear proper motion (result for the combined fit).\label{fig:1}}
\end{center}
\end{figure}

When data were fit using the two most stable and brightest maser spots (Spot ID:A and ID:B) together we derived an annual parallax of $\pi=0.213 \pm 0.026$ mas (\emph{uncertainty of} $12.4\%$), corresponding to a distance of $D = 4.69^{+0.65}_{-0.51}$ kpc. Results of the fitting procedure are illustrated in Fig.~\ref{fig:1}. When repeating the fitting procedure for these spots individually we arrived at consistent values for both the distance and error floors (\emph{see Table~\ref{table:2}}). It is clearly seen in Fig.~\ref{fig:1} that the data were fit well in the Dec. direction and poorly in the R.A. direction. The cause of this situation in the context of astrometric accuracy is revisited in the next subsection.

\citet{Mol96} were the first to estimate the kinematic distance to IRAS 20056+3350. Assuming that the source is at the near kinematic distance for Galactic rotation, and using the IAU recommended Galactic constants of R$_{0} = 8.5$ kpc and $\Theta_{0}=220$ km s$^{-1}$, they derived $D=1.67$ kpc from the radial velocity of of their NH$_{3}$ line observations at $v_{\rm LSR}=9.4$ km s$^{-1}$. To date, this value has been adopted in all subsequent literature regarding this source.
Assuming Galactic constants R$_{0}$ = 8.05 kpc and $\Theta_{0}$ = 235 km s$^{-1}$ (at V$_{\odot}$ = 15.3 km s$^{-1}$; \citealt{Honma12}) and the condition of flat Galactic rotation $\Theta$ = $\Theta_{0}$ = 235 km s$^{-1}$, we calculate the near and far kinematic distances of IRAS 20056+3350 as $D_{\rm near}$ = 1.2 kpc and $D_{\rm far}$ = 3.9 kpc respectively. Our distance obtained via trigonometric parallax is close to the far kinematic distance.

\null

\null

\begin{figure}[h!]
\begin{center}
\hspace{-0.4cm}
\includegraphics[scale=0.45]{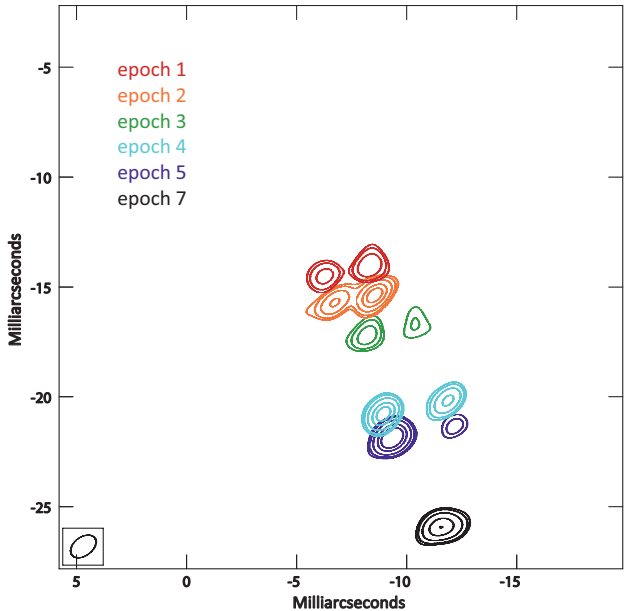}
\caption{Multi-epoch phase-referenced images of the prominent maser spots (\emph{Left} Spot ID:A and \emph{Right} ID:B). The background emission from noise has been removed artificially to highlight the emission of maser spots used in our analysis. Maser spots move from center to south-west, chronologically (epoch numbers 1 to 7 are shown, however epoch 6 was omitted for its proximity to epoch 5 in time, which confused the image - Spot ID:B was not detected in epoch 6). Contours shown are multiples of 3, 5, 10, 20 and 30 of the root mean squared noise for individual maps. The origin of the map is $(\alpha, \delta)_{\mathrm{J}2000.0}=(20^{\mathrm{h}}07^{\mathrm{m}}31^{\mathrm{s}}.2593$,
+33$^{\circ}$59'41".491), and the first epoch absolute co-ordinates of spot ID:A was $(\alpha, \delta)_{\mathrm{J}2000.0}=(20^{\mathrm{h}}07^{\mathrm{m}}31^{\mathrm{s}}.2586$,
+33$^{\circ}$59'41".477)
\label{fig:2}}
\end{center}
\end{figure}

\begin{figure}[h!]
\begin{center}
\includegraphics[scale=0.45]{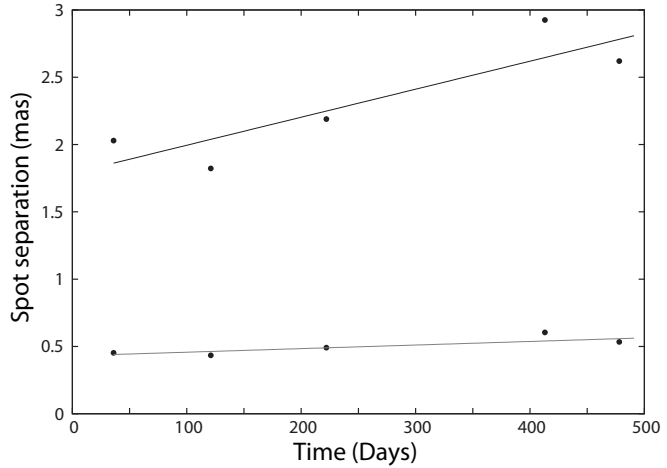}
\caption{Spatial separation of the two prominent maser spots (Spot ID:A and ID:B) in R.A. (above) and Dec. (below) over the first five epochs in which both maser spots were detected. In the absence of acceleration, the relative motions of maser spots should manifest as a linear increase or decrease in spot separation over time. Best fits to the spot separation in R.A. and Dec. are overlain. Standard deviations to the best fits in R.A. and Dec. directions were 0.19 mas and 0.04 mas respectively. The first side-lobe radius, as discussed in the text, appears near 2.5 mas in our data, and thus only affects the R.A. separation. \label{fig:3}}
\end{center}
\end{figure}

\subsection{Analysis of astrometric accuracy}
In the parallax fitting results of Fig.~\ref{fig:1}, the astrometric positions in the R.A. direction indicate a notably large degree of dispersion. Upon inspection of the brightness peaks in the emission maps, which we present in Fig.~\ref{fig:2}, emitting regions exhibit structural elongation primarily in the R.A. direction.
To investigate the astrometric error contribution from the structure of masers we monitored the spatial separation between the two bright, persistent maser spots. In the absence of acceleration, relative motions should be linear, thus we can compare the astrometric accuracy in R.A. and Dec. by evaluating the deviation from linear best fit motions in each respective direction. In Fig.~\ref{fig:3}, the large deviations from linearity seen in R.A. reveal an instability in the astrometric accuracy.
These deviations are likely caused by the notable elongation of maser structures, which results in an inaccurate determination of emission peaks in the interferrometric images. The perceived elongation may be real structure, or it may be an apparent effect caused by the smearing of the two spots at small separation from each other. Furthermore, the spatial separation of maser spots was often close to the first side-lobe radius in the R.A. direction (about 2.5 mas in our VERA observations). Interaction of the maser emission with side-lobes is likely to have contributed additional smearing of emission peaks in the R.A. direction.

The standard deviation values for the linear fits in R.A. and Dec. were 0.19 and 0.04 respectively. These values are similar to the error floors arrived at in the parallax fitting procedure which were required to reach a $\chi^{2}$ value of unity. Thus, our analysis suggests that the maser structure contributes a dominant source of astrometric error in these observations. Although this effect is detrimental for fitting the maser parallax in the R.A. direction, the astrometry in the Dec. direction is altogether unaffected, i.e this error is confined to the R.A. direction and thus supports preferential use of the Dec. direction offsets as the most suitable approach to parallax fitting.


\subsection{Proper motion of H$_{2}$O masers}

\begin{figure}[h!]
\begin{center}
\hspace{-0.3cm}
\includegraphics[scale=1.05]{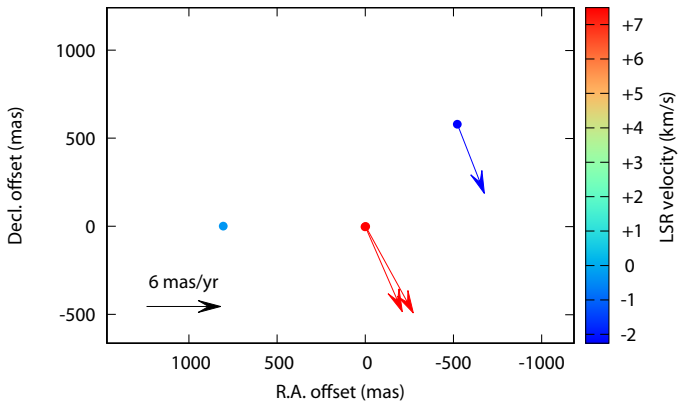}
\caption{Distribution and absolute sky-plane proper motions of H$_{2}$O maser spots in IRAS 20056+3350. At $D= 4.69$ kpc typical motions of $\sim$ 6 mas yr$^{-1}$ correspond to velocities of $\sim$ 135 km s$^{-1}$.
\label{fig:4}}
\end{center}
\end{figure}

We found a third maser spot (spot ID:C) in 3 consecutive epochs of the self-calibrated images. In order to deconvolve parallactic and secular motions a maser via fitting, a spot must be detected for around 1 year or longer in the phase-referenced images. However, we were able to estimate the proper motion of this spot via its relative motion to the bright maser spot (spot ID:A) detected in the self-calibrated images by assuming that both spots have a common annual parallax.
By combining the relative maser motions with the known motion of the reference maser we obtained the sky-plane proper motion of the third spot, where the error is calculated as the quadrature sum of the standard deviation in the relative motion of masers, and the proper motion error of the reference maser from parallax fitting. A fourth maser spot (spot ID:D) was detected in the final observation epoch only and thus its motion could not be found. Maser proper motions and their errors are listed in Table~\ref{table:2}. 
Absolute maser proper motions are shown in Fig.~\ref{fig:4} where arrows represent velocity vectors.

\vspace{-0.2cm}
No obvious spatial nor velocity structure can be inferred from our maser vector maps alone. With the small number of maser spots it is unwise to search our images for self-contained geometric structures arising from phenomena such as bipolar outflows and spherical expansions; such analysis is powerful only if the number of maser spots is large enough to reveal structure in the velocity field in regions of star formation (\emph{for example} \citealt{Imai11}).


\section{Discussion}

\subsection{The secular motion of IRAS 20056+3350}

\citet{Zhang05} presented an on-axis outflow mapped in CO which had a velocity centered at about $ +9$ km s$^{-1}$, in agreement with other molecular line observations (\emph{see Table 3}). The spectrum exhibits a triple-peak morphology with emission detected in the range of $0 \sim +20$ km s$^{-1}$. The peaks correspond to the blue-shifted outflow, the parent cloud and redshifted outflow, in ascending velocity. Since the $V_{\rm LSR}$ of our masers are all consistent with the blue limit of the CO emission we conclude that our observations are likely sampling masers associated with the blueshifted lobe of the outflow, which appears aligned to the line of sight. 
Although this does not allow direct interpolation of maser motions to a kinematic center, as is demonstrated in \citet{Imai11}, it does allow us to make the reasonable assumption that the sky-plane proper motions of maser spots with respect to the driving source are small, since the largest velocity components can be expected along the line of sight, in agreement with the direction of the molecular outflow.
As such, the average proper motion of maser spots should give a reasonable approximation to the secular motion of the SFR. 

\begin{figure}[h!]
\begin{center}
\includegraphics[scale=1.05]{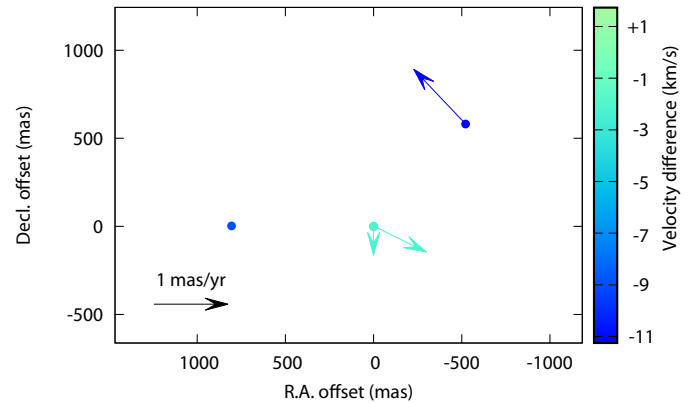}
\caption{Internal motions of H$_{2}$O maser spots in IRAS 20056+3350. The systemic motion is assumed to be the average proper motion of all measured maser spots on the sky, and the radial velocity of the ambient molecular gas, 9 km $s^{-1}$. At $D = 4.69$ kpc typical motions of $\sim$ 1 mas yr$^{-1}$ correspond to $\sim$ 22 km s$^{-1}$.
\label{fig:5}}
\end{center}
\end{figure}

We evaluated the group averaged proper motion for all maser spots in IRAS 20056+3350 to be 
($\mu_{\alpha}\cos\delta$, $\mu_{\delta}$) = ($-2.62\pm0.33$, $-5.65\pm0.52$) mas yr$^{-1}$.
Error values are the standard error of the mean, $\sigma$/$\sqrt{3}$, where $\sigma$ is the standard deviation of the proper motion of 3 spots.
After subtraction of the average group motion, residual motions reveal the internal kinematics of H$_{2}$O masers in IRAS 20056+3350. We present a map of the internal maser motions in Fig.~\ref{fig:5}.

The proper motions of SFRs can be used to break the distance ambiguity associated with kinematic distance estimates because sources at near and far kinematic distances are expected to exhibit different proper motions on the sky.
For IRAS 20056+3350 the near and far kinematic distances are $D_{\rm near}$ = 1.2 kpc and $D_{\rm far}$ = 3.9 kpc, respectively. Sources at these locations should show the proper motions of $\mu_{l} = -5.9$ mas yr$^{-1}$ and $\mu_{l} = -6.2$ mas yr$^{-1}$ respectively, if they rotate circularly around the Galactic center with $\Theta=220$ km s$^{-1}$ and assuming R$_{0}=8.05$ kpc \citep{Honma12}.

\begin{table*}[!t]
\caption{Systemic velocity, velocity width, and beam size of IRAS 20056+3350 observations.\label{table:3}}
\begin{center}
\small
\begin{tabular}{ccccccc}
\hline
Transition&$v_{\rm LSR}$&$\Delta v$&Beam size&\emph{Ref}&\\
 &$(km s^{-1})$&$(km s^{-1})$&(arcsec)&&\\ \hline
CS $J=(2 - 1)$&+8.8&3.2&39&\citet{Bronfman96}\\
CS $J=(5 - 4)$&+9.0&3.2&21&\citet{Jenness95}\\
C$^{18}$O $J=(2 - 1)$&+9.0&3.0&21&\citet{Jenness95}\\
$^{13}$CO $J=(1-0)$&+8.56&4.9&55&\cite{Wu01}\\
NH$_{3}$ $(1,1)$&+9.4&2.29&40&\citet{Mol96}\\
NH$_{3}$ $(2,2)$&+9.0&1.82&40&\citet{Mol96}\\
\hline
\end{tabular}
\end{center}
\end{table*}

\vspace{-0.2cm}
Using our estimate of the systemic proper motion of IRAS 20056+3350, ($\mu_{\alpha}\cos\delta$, $\mu_{\delta}$) = ($-2.62\pm0.33$, $-5.65\pm0.52$) mas yr$^{-1}$, we calculated the sky-plane motion of the source with respect to the Solar LSR in the direction of Galactic longitude as $\mu_{l} \cos b= -6.39 \pm 0.48$ mas yr$^{-1}$, using the standard solar motion (U$_{\odot}$,V$_{\odot}$,W$_{\odot}$) = (+10.3, +15.3, +7.7) km s$^{-1}$ (\citealt{Kerr86}, \emph{see also} \citealt{Ando11}). This proper motion value is consistent with that estimated for a source at the far kinematic distance. Furthermore, the far kinematic distance, $D_{\rm far}$ = 3.9 kpc, is closer to our distance of $D_\mathrm{tri}=4.96$ kpc, measured using annual parallax compared to the near distance $D_\mathrm{near}=1.2$ kpc

\subsection{The physical nature of IRAS 20056+3350}
\vspace{-0.1cm}
Since \citet{Mol96}, all astrophysical works regarding IRAS 20056+3350 adopt the near kinematic distance of 1.67 kpc in analysis of their data. Our trigonometric distance of $D = 4.69^{+0.65}_{-0.51}$ kpc is 2.8 times larger, thus significantly impacting the interpretation of data collected up to now. As such, we briefly revisit these past observations to summarise the nature of IRAS 20056+3350 - re-evaluated using the trigonometric distance and include also relevant distance-independent results to provide a full account of this MSFR. 


\null

\subsubsection{Interpreting archive data and using the revised distance}
\null
\vspace{-0.25cm}
\citet{CnW89} developed a diagnostic tool used to identify embedded massive OB type stars using IRAS point source catalogue colour criteria. Using their method IRAS 20056+3350 is a candidate for harbouring at least one embedded massive star.

The infrared luminosity of IRAS 20056+3350 from the total IRAS fluxes was estimated as L$_{\rm IRAS}=1100$L$_{\odot}$ \citep{Casoli86} using an assumed distance of 1.0 kpc. At the new distance, 4.69 kpc, it is revised as L$_{\rm IRAS}=24000$ L$_{\odot}$. The bolometric luminosity of IRAS 20056+3350 should be close to L$_\mathrm{IRAS}$, assuming that most of the emission from the embedded central star is absorbed and then re-emitted at infrared wavelengths. Using $M \propto L^{3.6}$ \citep{Allen76}, as was done in \citet{Casoli86}, the mass of a central star is revised from $ 7.0 M_\odot$ to 16.5 M$_{\odot}$, although the authors did not present the mass of the star. \citet{Casoli86} also produced a map in $^{13}$CO $J(1-0)$ with 4.'4 resolution using the Bordequx 2.5 m telescope and estimated a gas mass of M$_{\rm H2}= 56$ M$_{\odot}$ assuming local thermodynamic equilibrium (LTE), a distance of D $=1$ kpc and an abundance ratio of $^{13}$CO/H$_{2}$ of $2 \times 10^{-6}$ from \citet{Dick78}. The molecular gas mass is revised from 56 $M_\odot$ to  M$_{\rm H2}= 1200$ M$_{\odot}$ using the same assumption as that of \citet{Casoli86} but the revised distance.

Far infrared (FIR) continuum maps of IRAS 20056+3350 at 450 $\micron$ and 800 $\micron$ from \citet{Jenness95} clearly show dense cores, indicative of embedded star formation. 
Their Figures 1 and 2 show that the far infrared emission is associated with both the IRAS point source and water maser source, although the emission peak at $450 \mu m$ is offset from the masers by 5" in the direction of the cluster center. Although radio continuum emission at 8 GHz was marginally detected in their VLA observation, no compact source was identified (Figure 6 in \citealt{Jenness95}). This suggests that no H$_{\rm II}$ region is present, although \citet{Jenness95} do not comment explicitly on this.

To establish the radial velocity of the system we used published thermal molecular line observations since these typically trace ambient molecular gas. The systemic velocities, velocity widths, and telescope beam sizes from single-dish molecular line observations of IRAS 20056+3350 are summarised in Table.~\ref{table:3}. From these results we nominally conclude the LSR velocity of the system to be $+9.00 \pm 0.25$ km s$^{-1}$, where the error is the standard deviation of the listed values.

\begin{table*}[!t]
\caption{Flux parameters used in the SED fitting of IRAS 20056+3350.
\label{table:4}}
\begin{center}
\small
\begin{tabular}{cccccc}
\hline
Telescope &   Band   & Photometry & Aperture size & Source name & Mission reference \\
          & (name or $\mu$m) & (magnitude or flux) &   (arcsec)    &             &                   \\
\hline
UKIDSS    &    J     &  $18.985\pm0.105$ mag &      3       &J200731.43+335937.6&\citet{Lucas08}    \\
          &    H     &  $15.876\pm0.012$ mag &      3       &J200731.38+335939.4&                   \\
          &  K$_{s}$ &  $12.815\pm0.002$ mag &      3       &                   &                   \\
\hline
WISE     &    3.4     &  $7.961\pm0.024$ mag  &      8.5       &J200731.37+335940.9&\citet{Cutri14}    \\
         &    4.6     &  $4.780\pm0.04$ mag   &      8.5       &                   &                   \\
         &    12      &  $3.115\pm0.016$ mag  &      8.5       &                   &                   \\
         &    22      &  $-1.127\pm0.013$ mag &      16.5      &                   &                   \\
\hline
AKARI     &    9     &  $5510\pm59.2$ mJy  &      6       &2007315+335940  &\citet{Ishi10}    \\
          &   18     &  $9298\pm211$  mJy  &      6       &                &                  \\
\hline
IRAS     &    60     &  $422\pm63.3$   Jy  &      120       & 20056+3350 &\citet{IRAS88}    \\
         &    100    &  $757\pm121.12$ Jy  &      120       &            &                  \\
\hline
JCMT      &    450     &  $33\pm6.6$ Jy   &      3       &  20056+3350 &\citet{Jenness95}    \\
          &    800     &  $6.2\pm1.6$ Jy  &     12       &             &                     \\
\hline
\end{tabular}
\end{center}
\end{table*}

NH$_{3}$ (1,1) and (2,2) line observations of IRAS 20056+3350 were carried out by \citet{Mol96}. The measured line widths were $\Delta v_{(1,1)} = 2.29$ km s$^{-1}$ and $\Delta v_{(2,2)} = 1.83$ km s$^{-1}$, respectively. The NH$_{3}$ $(1,1)$ and $(2,2)$ lines predominantly trace the less dense envelope and the dense core, respectively. The authors argue that such sources, that exhibit lower velocity dispersion in the central core compared to the outer envelope, are characteristic of the \emph{lack} of an ultra compact H$_{\rm II}$ (UCHII) region, and possibly precede this stage.


\citet{Zhang05} mapped IRAS 20056+3350 in CO $J=(2-1)$ using the NRAO 12m telescope and reported a molecular outflow centered at the position of the IRAS source. Lobes of blueshifted and redshifted emission are concentric in the sky-plane indicating that the outflow is orientated to a pole-on geometry. Three velocity components are easily distinguishable in their spectra (panel ID 115 in Fig.2 of \citealt{Zhang05}). These components are symmetric about the center in the line profile: the blueshifted component which peaks at $+5$ km s$^{-1}$, the parent core at $+9$ km s$^{-1}$, and the redshifted component which peaks at $+13$ km s$^{-1}$. Emission was detected over the range of 0 to $+20$ km s$^{-1}$. From their CO $J=(2-1)$ data, \citet{Zhang05} estimated the outflow parameters of IRAS 20056+3350 using the diagnostics of \cite{Garden91}. Using the near kinematic distance, they estimated the mass, momentum and energy of the outflow as M$_{\rm outflow}= 2.3$ M$_{\odot}$, $p_\mathrm{outflow} = 20.6$ M$_{\odot}$ km s$^{-1}$ and E$_{\rm outflow} = 2.8 \times 10^{45}$ ergs. Our new distance gives outflow parameters M$_{\rm outflow}= 18$ M$_{\odot}$, $p_\mathrm{outflow}= 163$ M$_{\odot}$ km s$^{-1}$ and E$_{\rm outflow} = 2.2 \times 10^{46}$ ergs.

\subsubsection{Spectral energy distribution}

A practical method for investigating the physical nature and evolutionary stage of an astronomical object is to draw information from its spectral energy distribution (SED). In this effort we constructed the SED of IRAS 20056+3350 using published data.

Infrared fluxes were extracted from point source catalogues using the NASA Infrared Science Archive (IRSA), from the telescope missions of UKIDSS, WISE, AKARI, and IRAS. We also use JCMT fluxes at 450 and 800 $\mu$m from \citet{Jenness95}. 

\begin{figure}[h!]
\begin{center}
\vspace{-0.4cm}
\includegraphics[scale=0.35]{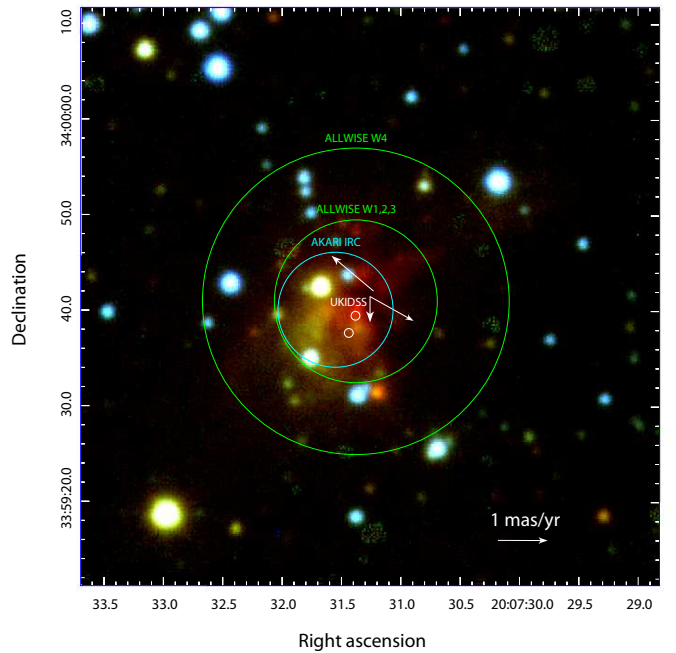}
\caption{Aperture size and positions taken from point source catalogues of various infrared telescope missions. These correspond to photometric measurements used to evaluate the SED of IRAS 20056+3350, as is discussed in the main text. Apertures are plotted over the near-IR composite image from UKIDSS where colours \emph{blue, green} and \emph{red} correspond to bands J, H and K$_{s}$. Proper motions of H$_{2}$O masers observed with VERA are indicated by white arrow vectors. The IRAS aperture was too large to be displayed in this figure.\label{fig:6}}
\end{center}
\end{figure}

The aperture sizes or angular resolutions of each of the catalogs were different. 
In Figure.~\ref{fig:6} we show a three-colour image of IRAS 20056+3350 made using published data from UKIDSS in J, H, K$_{s}$ bands, and whose catalogue has the the highest angular resolution of any we used. A distinctly redder region is seen in the center-western portion of the cluster. A star forming core with redder colours has more intense radiation at the longer wavelengths which is characteristic of cores which are forming more massive stars. 
We identify this region as the most likely site of embedded massive stars.
Furthermore, a dark lane which is prominent in the $K$-band \citep{Ver10} and the association of H$_{2}$O masers support that this is a site of embedded star formation.
Thus, we required that flux apertures from all point source catalogues be positionally consistent with this region. Details of the fluxes used in SED fitting are given in Table~\ref{table:4} and the position and sizes of apertures are overlaid onto Figure.~\ref{fig:6}.

Although the beams of the larger apertures will invariably contain multiple cluster members, we made no corrections for this because our target should be the brightest in the beam; our target is the most massive and the flux from other objects should be negligible, since the luminosity is extremely sensitive to the mass of the object, by the relation L $\propto$ M$^{3.6}$. 

To test the reliability of using the IRAS 60 and 100 $\micron$ bands in our SED we compared the photometries of IRAS $F_{24\micron}$ and WISE $F_{22\micron}$ and found them to be consistent, supporting the reliability of using the IRAS fluxes.
450 and 800 $\micron$ fluxes, measured using the James Clarke Maxwell telescope (JCMT) by \citet{Jenness95}, were evaluated with an aperture equivalent to the extent of the FWHM of the observed emission. Inclusion of these fluxes may lead to an overestimation of the integrated SED flux since hot gas associated with the outflow will be included. However, since the core emission is compact and centered on the maser source, the flux is likely dominated by the embedded massive star. These data provide valuable confines for the longer wavelength portion of the SED.

\begin{figure}[h!]
\begin{center}
\hspace{-0.4cm}
\includegraphics[scale=0.75]{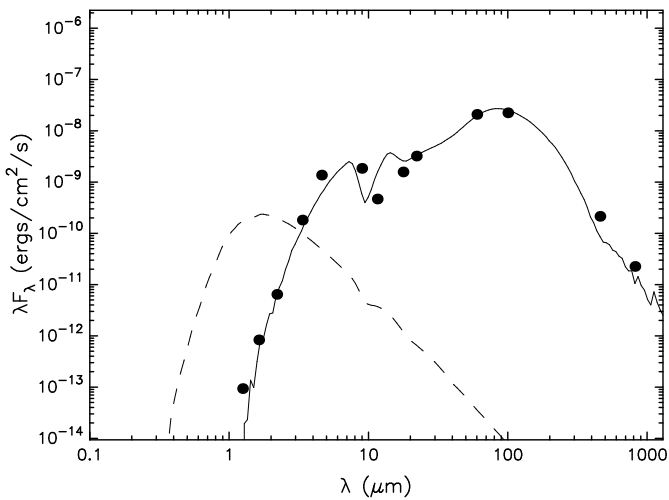}
\caption{SED of IRAS 20056+3350 using data from UKIDSS, AKARI, WISE, IRAS and JCMT.
\label{fig:7}}
\end{center}
\end{figure}

To interpret our photometric data we use the fitting tool of \cite{Rob07} which is well documented in the introductory paper referenced here. The SED of IRAS20056+3350, according to the model fitting is shown in Fig.~\ref{fig:7}.

The SED model corresponds to a 18.4 M$_{\odot}$ central star of age $10^{4}$yrs, embedded in an envelope of mass 3300 M$_{\odot}$. The total luminosity of the region is $2.45\times 10^{4} L_{\odot}$. 
The luminosity and stellar mass derived from the SED model agrees well with the estimates made using the four IRAS bands by of \citet{Casoli86} re-evaluated using our distance estimate as $L_{\rm IRAS}=24000 L_{\odot}$ and $ M_{*}\sim 16.5 M_{\odot}$. Furthermore, the envelope mass derived from the SED model agrees within a factor of two with the estimate of the molecular hydrogen content made by \citet{Casoli86} re-evaluated using our distance estimate as $M_{H_{2}}=1200 M_{\odot}$.

Interestingly, the inclination angle required to produce the best-fit model is 18$^{\circ}$ to the line of sight, which agrees with the general orientation inferred from the CO $J=(2-1)$ observations of \citet{Zhang05}. The ability of the SED fitting software to determine the inclination of young stellar object (YSO) systems solely from photometry data has been demonstrated for G34.4+0.23MM in \citet{Rob08} and for IRAS 04368+2557 in \citet{Rob07}. In these cases the system inclination had been previously established observationally from the orientation of bipolar outflows and was well matched by the SED fitting tool.

It should be pointed out that the successful fit of the SED model to the photometry data shows only that there exists a theoretical model \emph{consistent} with the data and it is not \emph{proof} that the model results truly reflect the nature of the source. The parameters directly determined from the data are luminosity and temperature of the emitting region, the further details (stellar mass, envelope mass, inclination, etc.) depend on the evolutionary tracks applied to the data, this is cautioned in \citet{Rob08}. However, in our case the SED results are consistent with the picture drawn from an abundance of data from past observations re-evaluated at the trigonometric distance. It is this consistency rather than the pleasing appearance of the model fit to the data, that raises confidence in our results. These consistencies, along with the common proposed inclination angle, supports the basis of our argument; that the secular proper motion of IRAS 20056+3350 can be reasonably estimated from the group motions of H$_{2}$O masers associated with the line of sight outflow.

We conclude that IRAS 20056+3350 is a distant site of massive star formation at a young stage in evolution. The existence of an embedded, young massive star is strongly suggested by the IRAS colours, SED, and presence a compact sub-millimeter core. With respect to youth, the evolutionary stage precedes the formation of an ultra compact H$_{\rm II}$ (UCHII) region, this is suggested from the turbulence ratio of the core and envelope seen in NH$_{3}$ and is corroborated by the lack of significant centimeter structure at 8GHz.
Line spectra of core tracing molecular transitions such as CS reveal substantial turbulence (\emph{see} Table.~\ref{table:3}) which is suggested by \citet{Ao04} to be characteristic of young SFRs which have not yet dissipated primordial turbulent motions inherent in the core.
An active bipolar molecular outflow, seen in spectral wings and emission map, appears aligned to the line of sight - which we believe to be associated with the H$_{2}$O masers we observed with VERA due to the consistency of velocities.
The overall picture ascertained from past observations is corroborated by the SED of IRAS20056+3350, which resembles a YSO in the Class I evolutionary phase.


\null

\subsection{Structure of the Local Arm}

Full-scale studies of the structures of the spiral arms have recently been made available from analysis of the joint results of VLBI annual parallax measurements of MSFRs. Recent cases are; the Local Arm by \citet{Xu13}, the Perseus Arm by \citep{Zhang13,Choi14}, the Saggitarius Arm by \citet{Wu14} and the general structure is discussed in \citet{Reid14}. IRAS 20056+3350 is the most distant MSFR in the Local arm for which a trigonometric distance has been determined. Therefore, although we cannot justify a large-scale re-evaluation of the Local arm, we can provide an important contribution to the picture recently provided by the aforementioned authors.
The nature of the Local arm is discussed particularly in the work of \citet{Xu13} where the authors consider the three possible scenarios; \emph{1.} That the Local Arm is a branch of the Perseus Arm. \emph{2.} That the Local Arm may be part of a major arm and connects to the Carina Arm. \emph{3.} That it is an independent arm segment - a `spur'.

\begin{figure}[h!]
\begin{center}
\hspace{-0.6cm}
\includegraphics[scale=0.182]{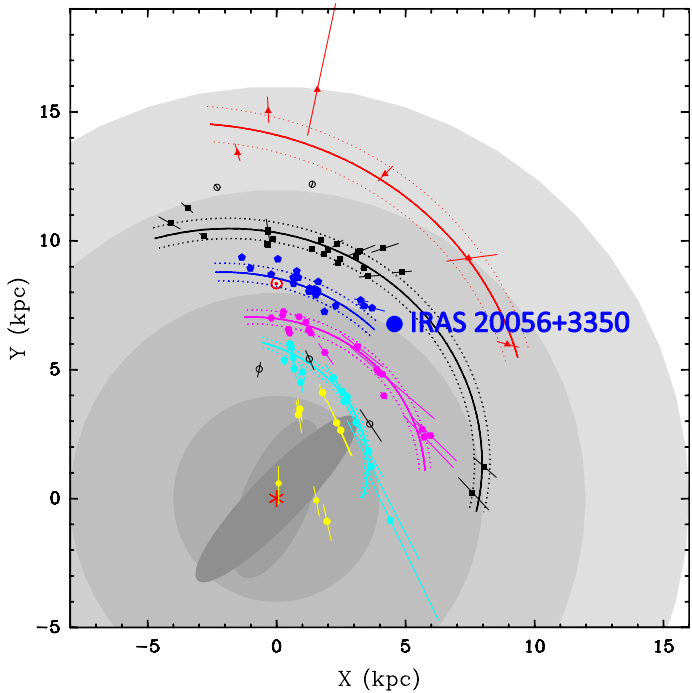}
\caption{Location of IRAS 20056+3350 in relation to the Galactic spiral structure as evaluated using VLBI observations of MSFRs, compiled by \citet{Reid14} (\emph{individual source references therein}).
\label{fig:8}}
\end{center}
\end{figure}

In Fig.~\ref{fig:8} we show the position of IRAS 20056+3350 in the context of the Galactic spiral structure, which was recently compiled by \citet{Reid14} (\emph{References for individual SFRs therein}). In drawing logarithmic spiral arms they used the Galactocentric distance of the Sun, R$_{0}=8.34$ kpc.

\null

As seen in Fig.~\ref{fig:8}, the Galactic location of IRAS 20056+3350 diverges somewhat from the logarithmic curve determined for the Local arm  by \citet{Reid14}. Deviation is in the direction of the Perseus arm and follows the example of three other MSFRs at the leading tip of the arm; G075.76+00.33 \citep{Xu13}, G075.78+00.34 \citep{Ando11,Xu13} and AFGL 2591 \citep{Rygl12}. The location of IRAS 20056+3350 is generally consistent with what would be expected in the scenario where the Local Arm joins onto the Perseus arm further down into the spiral pattern - scenario \emph{1.} of \citet{Xu13}. However, the alternative scenarios cannot be ruled out until this region of the Galaxy is mapped in more detail.

\null

\subsection{Evaluation of the $\Omega_{0}$ Galactic constant}

From our trigonometric distance we calculated the Galactocentric distance of IRAS 20056+3350 to be $R_{*}=7.91$ kpc, when using R$_{0}$ = 8.05 kpc from \citet{Honma12}. Since $R_{*} \simeq R_{0}$, and from the low value of $v_{\rm LSR} = 9$ km s$^{-1}$, it is evident that IRAS 20056+3350 resides spatially and kinematically near to the Solar circle.

\cite{Nagayama11a}  demonstrated that the angular velocity of Galactic rotation at the Galactocentric distance of the Sun, $\Omega_{0}$, can be estimated for objects near the Solar circle and tangent points, with relaxed assumptions on the adopted value of $R_{0}$.  For objects with negligible peculiar motion, the angular velocity of Galactic rotation at the Sun is given by

\begin{eqnarray}
\Omega_{0} &=& -   a_{0}\mu_{l}+  v_{r}\left(\frac{1}{D\tan l}-\frac{1}{R_{0}\sin l}\right)\label{eq:3}
\end{eqnarray}
\null

\noindent where $D$ is the distance to the source from the Sun, $v_r$ is the LSR velocity, $\mu_l$ is the systemic sky-plane proper motion in the direction of Galactic longitude, and $a_0=4.74$ km s$^{-1}$ kpc$^{-1}$ (mas yr$^{-1}$)$^{-1}$ is a unit conversion factor. 

In order to make use of Equation (1) we first convert our systematic proper motion estimate into the proper motion with respect to the LSR in the Galactic coordinate system. To correct for the motion of the Sun with respect to the LSR we use the following values: (U$_{\odot}$,V$_{\odot}$,W$_{\odot}$) = (+10.3, +15.3, +7.7) km s$^{-1}$ (\citealt{Kerr86}, \emph{see also} \citealt{Ando11}).
The relevant parameters, determined in this paper, for IRAS 20056+3350 are:
$D = 4.69^{+0.65}_{-0.51}$ kpc and $\mu_{l} = -6.39 \pm 0.48$ mas yr$^{-1}$. We use $v_r = +9.0 \pm 0.25$ km s$^{-1}$ determined from molecular line observations (\emph{see section 4.2.1}). Applying the above parameters to Equation (1), and using R$_{0}$ = 8.05 kpc, gives \\

$\Omega_{0} = 29.76\pm2.29$ km s$^{-1}$ kpc$^{-1}$\\

\noindent We demonstrate the invariability of this approach on the adopted value of R$_{0}$ for Solar circle objects by recalculating $\Omega_{0}$ for a range of values of $7 \le R_{0} \le 9$ kpc. This difference produces only a small deviation of $\Delta \Omega_{0} = \pm 0.3$ km s$^{-1}$ kpc$^{-1}$. The value of $\Omega_{0}$ is consistent with other Solar circle and tangent point sources as discussed in detail in \citet{Burns14}. All such objects produce a value that is higher than that obtained from the ratio of the Galactic constants recommended by the IAU of $\Omega_{0} =\Theta_{0} / R_{0} = 25.9$ km s$^{-1}$ kpc$^{-1}$ \citep{Kerr86}.

\section{Conclusions}

\noindent We measured the trigonometric distance of IRAS 20056+3350 to be $D = 4.69^{+0.65}_{-0.51}$ kpc, which is 2.8 times larger than the near $D_{kin}$ often adopted in the literature. In measuring the annual parallax the astrometric accuracy in the R.A. and Dec. directions were investigated and the former was found to be significantly influenced by the elongation of maser structure, although the latter was affected negligibly. The astrometric errors evaluated in this analysis were well matched by the error floors required to produce a $\chi^{2}$ value of unity in the parallax fitting procedure.\\

\noindent We determined the systematic proper motion of the source to be ($\mu_{\alpha}\cos\delta$, $\mu_{\delta}$) = ($-2.62\pm0.33$, $-5.65\pm0.52$) mas yr$^{-1}$ under the assumption that the H$_{2}$O masers used in our observations are associated with a line of sight bipolar outflow. This assumption is justified by the spatial and kinematic proximity of masers to the blue lobe of the bipolar outflow as revealed from archive maps, velocity spectra and an SED model compiled using infrared photometry data interpreted using the fitting tool of \citet{Rob07}.\\

\noindent From the accumulation, re-evaluation and summary of various archive data we find IRAS 20056+3350 to be a young MSFR which is forming a central object of considerable mass ($>$16$M_{\odot}$). It is at an evolutionary stage resembling a \emph{Class I} YSO and appears to precede the formation of an UCHII region.\\ 

\noindent IRAS 20056+3350 is at the leading tip of the Local arm, near the Solar circle, and its position in relation to the spiral arms tentatively supports that the Local arm may be a branch of the Perseus arm.\\

\noindent Using our results obtained with VERA, and the special geometry applicable to Solar circle objects, we evaluated the angular velocity of Galactic rotation at the Sun,
$\Omega_{0} = 29.75\pm2.29$ km s$^{-1}$ kpc$^{-1}$, which is consistent with other Solar circle and tangent point sources estimated using the same procedure. IRAS 20056+3350 included, all such estimates are higher than the value derived from the ratio of the Galactic constants recommended by the IAU of $\Omega_{0}=\Theta_{0} / R_{0} = 25.9$ km s$^{-1}$ kpc$^{-1}$.

\null

\null

This research made extensive use of NASA's Astrophysics Data System (ADS), the SIMBAD database and VizieR catalogue access tool, operated at CDS, Strasbourg, France, and the NASA/IPAC Infrared Science Archive, which is operated by the Jet Propulsion Laboratory, California Institute of Technology, under contract with the National Aeronautics and Space Administration

R.B. would like to thank Andrej Sobolev for stimulating and encouraging discussion on this SFR.

R.B. would like to acknowledge the Ministry of Education, Culture, Sports, Science and Technology (MEXT), Japan for financial support under the Monbukagakusho scholarship.

\null
\begin{small}

\bibliographystyle{apj}
\bibliography{I20056}
\end{small}

\end{document}